\input{aipcheck}

\documentclass[
    ,final            
  ]
  {aipproc}

\layoutstyle{6x9}

\begin{document}

\title{Search for Proton Medium Modifications in the $^4$He$(\vec e,e' \vec p)^3$H Reaction}


\author{S. Strauch for the Jefferson Lab Hall A Collaboration}{
  address={University of South Carolina, Columbia, South Carolina 29208, USA}
}

\begin{abstract}
  Polarization transfer in quasi-elastic nucleon knockout is sensitive
  to the properties of the nucleon in the nuclear medium, including
  possible modification of the nucleon form factor and/or spinor.  In
  our recently completed experiment E03-104 at Jefferson Lab we
  measured the proton recoil polarization in the $^4$He($\vec
  e,e^\prime \vec p\,$)$^3$H reaction at a $Q^2$ of 0.8 (GeV/$c$)$^2$
  and 1.3 (GeV/$c$)$^2$ with unprecedented precision. These data
  complement earlier data between 0.4 and 2.6 (GeV/$c$)$^2$ from both
  Mainz and Jefferson Lab. The measured ratio of polarization-transfer
  coefficients differs from a fully relativistic calculation, favoring
  either the inclusion of a medium modification of the proton form
  factors predicted by a quark-meson coupling model or strong
  charge-exchange final-state interactions. The measured induced
  polarizations agree well with the fully relativistic calculation and
  indicate that these strong final-state interactions may not be
  applicable.
\end{abstract}

\maketitle


\section{Introduction}

The observation of neutrinos involves the measurement of secondary
particles they create when interacting with nuclear targets. The
analysis of these data requires thus a reliable understanding of the
neutrino-nucleus reaction; see {\it e.g.}  the calculations by
Szczerbinska {\it et al.} \cite{Szczerbinska07} and Benhar {\it et
  al.} \cite{Benhar05}. The advent of high-intensity neutrino beams
makes neutrino experiments with increased statistical precision
possible and put strong demands on the reduction of systematic
uncertainties related to the modeling of neutrino interactions.

Relativity and final-state interactions (FSI) are two essential model
ingredients for a successful description of the electroweak cross
sections \cite{Amaro07}.  A model ingredient which is not often
considered are possible modifications of the internal structure of
nucleons bound in the nuclear medium. However, a calculation by Lu
{\it et al.}~\cite{Lu98}, using a quark-meson coupling (QMC) model,
suggests a measurable deviation from the free-space electromagnetic
form-factor over the $Q^2$ range 0.0 $< Q^2 <$ 2.5
(GeV/$c$)$^2$. Similar measurable effects have recently been
calculated in a light-front-constituent quark model by Frank {\it et
  al.} \cite{Fr96}, a modified Skyrme model by Yakshiev {\it et
  al.}~\cite{Ya02}, a chiral quark-soliton model by Smith and Miller
\cite{Smith04}, and the Nambu-Jona-Lasinio model of Horikawa and Bentz
\cite{Horikawa05}.  The connection between the modifications induced
by the nuclear medium of the nucleon form factors and of the deep
inelastic structure functions is discussed by Liuti \cite{Liuti06}
using the concept of generalized parton distributions.

Tsushima, Kim, and Saito have applied the QMC model to calculate the
effect of bound nucleon form factors on charged-current
neutrino-nucleus scattering \cite{Tsushima04}. They estimate the
effect on the inclusive $^{12}$C$(\nu_\mu,\mu^-)X$ cross section and
find a reduction of $\approx 8$\% for the total cross section. The
correction due to the in-medium form factors could indeed be
significant for a precise estimate of the charged-current
neutrino-nucleus cross section.

Electron-nucleus scattering data provide strong constraints on nuclear
modeling of neutrino reactions. The polarization-transfer in
quasielastic nucleon knockout, $(\vec e, e^\prime \vec N)$, is sensitive to
the properties of the nucleon in the nuclear medium, including
possible modification of the nucleon form factor and/or spinor. This
reaction is also believed to have minimal sensitivity to
reaction-mechanism effects; see, {\it e.g.}, \cite{Laget94}.  In
free electron-nucleon scattering, the ratio of the electric to
magnetic Sachs form factors, $G_E$ and $G_M$, is given by \cite{Ar81}:
\begin{equation}
\frac{G_E}{G_M} = -\frac{P'_x}{P'_z} \cdot \frac{E_e + E_{e'}}{2m_p}
\tan \left(\frac{\theta_e}{2} \right).
\label{eq:free}
\end{equation}
Here, $P'_x$ and $P'_z$ are the transferred polarizations, transverse
and longitudinal to the proton momentum. The beam energy is $E_e$, the
energy (angle) of the scattered electron is $E_{e'}$ ($\theta_e$), and
$m_p$ is the proton mass. 

In the following, the results of $^4$He$(\vec e,e^\prime \vec p)^3$H
experiments to study the proton knock-out process will be
discussed. In these experiments $^4$He was chosen as target because of
its high density and relatively simple structure, which facilitate
microscopic calculations.

\section{Experiments}

The first $^4$He$(e,e^\prime \vec p)^3$H polarization-transfer measurements
were performed at the Mainz microtron (MAMI) at a four-momentum
squared $Q^2 = 0.4$ (GeV/$c$)$^2$ \cite{mainz4he} and as experiment
E93-049 at Jefferson Lab Hall A at $Q^2$ = 0.5, 1.0, 1.6, and 2.6
(GeV/$c$)$^2$ \cite{strauch03}. Our recent experiment E03-104
\cite{e03104} has extended these measurements with two high-precision
points at $Q^2$ = 0.8 and 1.3 (GeV/$c$)$^2$. The data were taken in
quasielastic kinematics at low missing momentum with symmetry about
the three-momentum-transfer direction to minimize conventional
many-body effects in the reaction.  In each experiment, two
high-resolution spectrometers were used to detect the scattered
electron and the recoiling proton. The missing-mass technique was
employed to identify $^3$H in the final state.  The
proton recoil-polarization observables were extracted by means of the
maximum-likelihood technique, utilizing the azimuthal distribution of
protons scattered by the graphite analyzer in a focal-plane
polarimeter (FPP) \cite{Punjabi05}.

Since these experiments were designed to detect differences between
the in-medium polarizations compared with the free values, both $^4$He
and $^1$H targets were used. The results of the measurements are
expressed in terms of the polarization-transfer double ratio, in which
the helium polarization ratio is normalized to the hydrogen
polarization ratio measured in the identical setting:
\begin{equation}
R = \frac{(P_x'/P_z')_{^4\rm He}}{(P_x'/P_z')_{^1\rm H}}.
\label{eq:rexp}
\end{equation}
As a cross-check, the hydrogen results were also used to extract the
free-proton form-factor ratio $G_E/G_M$, which was found to be in
excellent agreement with previous polarization-transfer data
\cite{Punjabi05,gep}.  Nearly all systematic uncertainties cancel in $R$: the
polarization-transfer observables are to first order independent of
instrumental asymmetries in the FPP, and their ratio is independent of
the electron-beam polarization and carbon analyzing power. Small
systematic uncertainties are predominantly due to uncertainties in the
spin transport through the proton spectrometer.

The induced proton polarization $P_y$ in $(e,e^\prime \vec p)$ is a direct
measure of final-state interactions and was also measured in the Jefferson Lab
experiments. Its extraction is complicated by the presence of instrumental
asymmetries.

\section{Results}

\begin{figure}[htb!]
  \includegraphics[width=\textwidth]{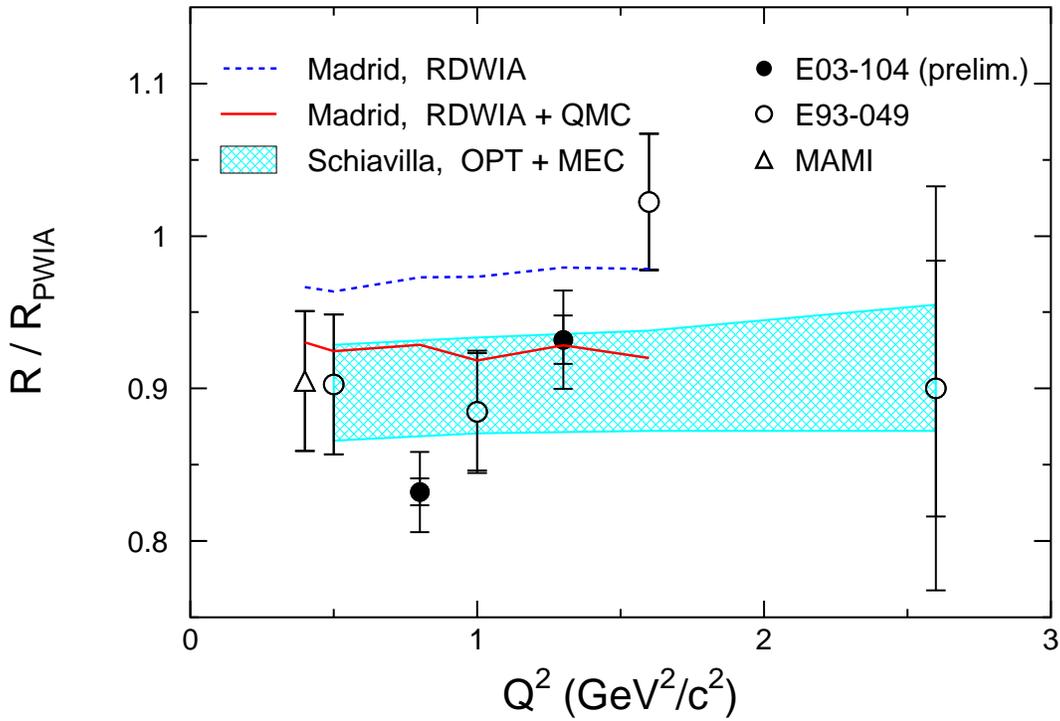}
  \caption{Superratio
    $R/R_{\rm PWIA}$ as a function of $Q^2$  from Mainz \cite{mainz4he} and
    Jefferson Lab experiment E93-049 \cite{strauch03} (open symbols) along with
    preliminary results from experiment E03-104 (filled circles); $R$
    is the ratio of transverse to longitudinal polarization of the
    recoiling proton in $^4$He$(\vec e,e^\prime \vec p)^3$H compared to the
    same ratio for $^1$H$(\vec e, e^\prime \vec p)$. The baseline
    $R_{\rm PWIA}$ is the value of $R$ obtained in a plane-wave
    calculation, to account for the 'trivial' effects of free {\it vs.} moving
    proton. The data are compared
    to calculations from the Madrid group \cite{Ud98} and Schiavilla
    {\it et al.} \cite{Schiavilla05}. 
    \label{fig:ratioplot}}
\end{figure}

The polarization-transfer double ratio, $R$, is shown in
Figure~\ref{fig:ratioplot}.  The preliminary data of the present
experiment E03-104 (filled circles) are consistent with our previous
data from E93-049 \cite{strauch03} and MAMI \cite{mainz4he} (open
symbols). The final systematic uncertainties from E03-104 are expected
to be much smaller than those shown here. The data are compared with
relativistic distorted-wave impulse approximation (RDWIA) calculations
from the Madrid group \cite{Ud98}. The RDWIA calculation (dashed curve)
overpredicts the data by about 6\%. After including the density
dependent medium-modified form factors as predicted by Lu {\it et al.}
\cite{Lu98} in the RDWIA calculation (solid curve) good
agreement with the previous data is obtained.\footnote{A Glauber-based
calculations by Ryckebusch \cite{Lava05} gives $R/R_{\rm PWIA}$ of
about one and is not shown here. We also do not show a chiral-soliton
model calculation of the in-medium form factors by Smith and Miller
\cite{Smith04}, which give results similar to the QMC results.}
This agreement has been interpreted as a possible evidence of proton
medium modifications \cite{strauch03}. This interpretation
is based on the more effective description of the data in a particular 
model in terms of medium modifications of nucleon form factors and 
requires to have
excellent control of the inherent many-body effects, such as
meson-exchange currents (MEC) and isobar configurations (IC). In
addition, when probing nuclear structure, one has to deal with
final-state interactions (FSI). Indeed, the observed suppression of
the polarization-transfer ratio has been equally well described in a
more traditional calculation by Schiavilla \cite{Schiavilla05} using
free form factors by including charge-exchange final-state
interactions and two-body charge and current operators (shaded band).
Yet, the choice of the parameters for the charge-exchange FSI is not well
constrained by data.

The preliminary results from E03-104 possibly hint at an unexpected
trend in the $Q^2$ dependence of $R$. This is particularly interesting
as other calculations of in-medium form factors suggest a different
$Q^2$ dependence than that of the QMC model. 
The high statistics of the data from E03-104 will allow
for a study of the individual polarization observables as a function
of missing momentum up to about 120 MeV/$c$.

\begin{figure}
  \includegraphics[width=\textwidth]{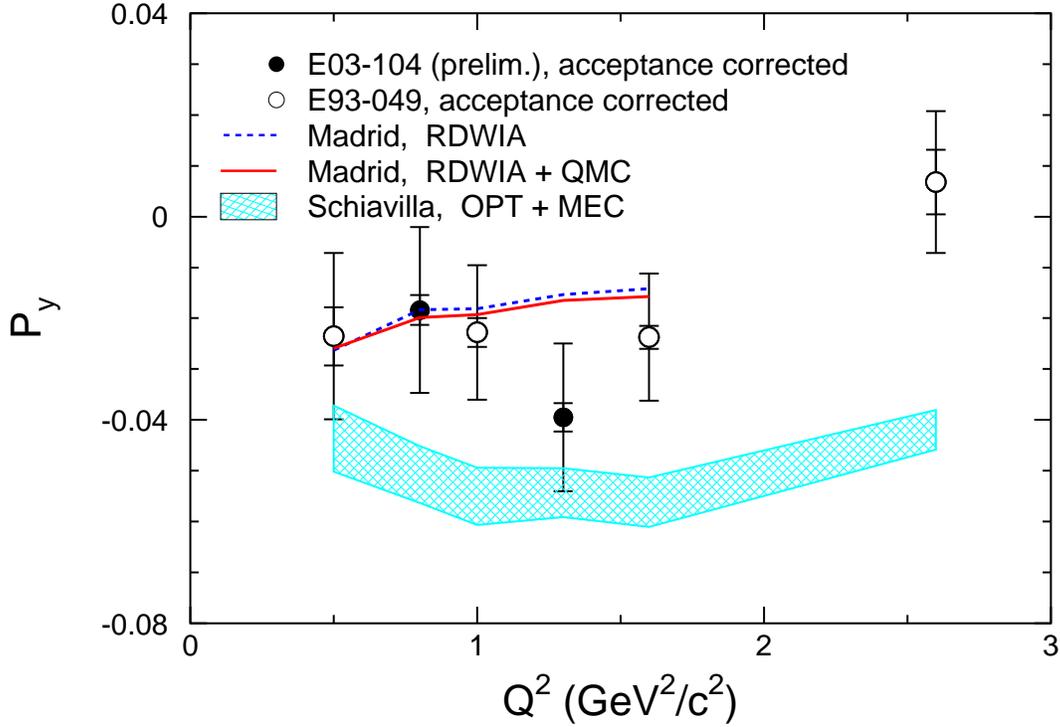}
  \caption{Induced polarization data from Mainz \cite{mainz4he} and
    Jefferson Lab experiment E93-049 \cite{strauch03} along with preliminary
    results from experiment E03-104. The data are compared to
    calculations from the Madrid group \cite{Ud98} and Schiavilla {\it et al.}
    \cite{Schiavilla05}. The comparison is made for missing momentum
    $p_m \approx 0$; note that the experimental data have been
    corrected for the spectrometer acceptance for this comparison.  }
  \label{fig:py}
\end{figure}

The difference in the modeling of final-state interactions is the
origin of the major part of the difference between the results of the
calculations by Udias {\it et al.} \cite{Ud98} and Schiavilla {\it et
  al.}  \cite{Schiavilla05} for the polarization observables. Effects
from final-state interactions can be studied experimentally with the
induced polarization, $P_y$.  Figure \ref{fig:py} shows the data for
$P_y$. The induced polarization is small in this reaction. The sizable
systematic uncertainties are due to possible instrumental asymmetries.
Dedicated data have been taken during E03-104 to study these and we
hope to significantly reduce the systematic uncertainties in $P_y$ in
the final analysis. The data are compared with the results of the
calculations from the Madrid group and Schiavilla {\it et al.} at
missing momenta of about zero. The data have been corrected for the
spectrometer acceptance to facilitate this comparison.
Presently, the data suggest that the magnitude of the induced
polarization (and thus the final-state interaction) is overestimated
in the model of Schiavilla {\it et al.}.  A comparison of the model
calculations in Figure \ref{fig:ratioplot} and Figure \ref{fig:py}
shows that the in-medium form factors (solid curves) mostly affect the
ratio of polarization-transfer observables, not the induced
polarization. It is a key element of E03-104 to have access to both.
 
In summary, polarization transfer in the quasielastic $(e,e^\prime p)$ reaction
is sensitive to possible medium modifications of the bound-nucleon
form factor, while at the same time largely insensitive to other
reaction mechanisms. Currently, the $^4$He$(\vec e, e^\prime \vec p)^3$H
polarization-transfer data can be well described by either the
inclusion of medium-modified form factors or strong charge-exchange
FSI in the models. However, these strong FSI effects may not be
consistent with the induced polarization data. The final analysis of
our new high-precision data from Jefferson Lab Hall A should provide a more
stringent test of these calculations.


\begin{theacknowledgments}
This work was supported in parts by the U.S. National Science Foundation: NSF PHY-0555604.
Jefferson Science Associates operates the Thomas Jefferson National Accelerator Facility under DOE
contract DE-AC05-06OR23177.
\end{theacknowledgments}

\end{document}